\documentclass[conference]{IEEEtran}
\IEEEoverridecommandlockouts
\usepackage[utf8]{inputenc}
\usepackage{blindtext, graphicx}
\usepackage[utf8]{inputenc}
\usepackage[english]{babel}
\usepackage{amsmath,amssymb,amsfonts}
\usepackage{mathtools}
\usepackage{caption}
\usepackage{subcaption}
\usepackage{pgffor}
\usepackage{enumerate}
\let\labelindent\relax
\usepackage{enumitem} 
\usepackage{babel}
\usepackage{chemgreek,textgreek}
\usepackage{tikz}
\usepackage{pifont}
\newcommand{\cmark}{\ding{51}}%
\newcommand{\xmark}{\ding{55}}%
\usepackage{tikz}
\usepackage{enumitem}
\usepackage{graphicx,wrapfig,lipsum}
\usepackage{amsmath,hyperref}
\usepackage{multirow}
\usepackage{multicol}
\usepackage{todonotes}
\usepackage{xcolor}
\usepackage[format=hang]{caption}
\usepackage{wrapfig}

\usepackage{todonotes}
\ifCLASSINFOpdf
\else
\fi

\begin{document}
%
\title{empathi: An ontology for Emergency Managing and
Planning about Hazard Crisis }


\author{
\authorblockN{Manas Gaur\authorrefmark{2}, 
Saeedeh Shekarpour\authorrefmark{1}, 
Amelie Gyrard\authorrefmark{2}, 
Amit Sheth\authorrefmark{2}}
\IEEEauthorblockA{\authorrefmark{1} University of Dayton, USA, 
 sshekarpour1@udayton.edu}
\IEEEauthorblockA{\authorrefmark{2} Kno.e.sis Center,  \{manas,amelie,amit\}@knoesis.org}
}


%


\maketitle

\begin{abstract}
In the domain of emergency management during hazard crises, having sufficient situational awareness information is critical.
It requires capturing and integrating information from sources such as satellite images, local sensors and social media content generated by local people.
A bold obstacle to capturing, representing and integrating such heterogeneous and diverse information is lack of a proper ontology which properly conceptualizes this domain, aggregates and unifies datasets.
Thus, in this paper, we introduce \emph{empathi} ontology which conceptualizes the core concepts concerning with the domain of emergency managing and planning of hazard crises.
Although \emph{empathi} has a coarse-grained view, it considers the necessary concepts and relations being essential in this domain.
This ontology is available at \url{https://w3id.org/empathi/}.
\end{abstract}



%
\IEEEpeerreviewmaketitle

\section{Introduction}

We can not change the nature. However, we can promote our planning, preparation, and response
strategies about crises happening in the three phases of hazards, i.e. (i) pre-hazard, (ii) in-hazard and (iii) post-hazard.
Currently, a substantial body of situational information collected from 
sources such as satellite images, sensors, social media content generated by people who are involved in crisis reporting and response, etc.
Indeed, harnessing and exploiting this \emph{ hazard-related Big Data} is an essential means towards situational awareness which helps to manage crises, threads and risks of hazards in each phase.
Despite the availability of such data, still there is a significant deficiency in \textbf{representing}, \textbf{annotating} and more importantly \textbf{integrating} this heterogeneous hazard-related big data.
This deficiency can be relieved by providing an ontology which \emph{conceptualizes} and \emph{organizes} situational and environmental awareness data (events, activities) subjected to hazards of any kind.
Our investigation in the state-of-the-art hazard-related conceptualization (i.e., taxonomy, vocabulary and ontology) revealed existing of a few works which mainly conceptualized either hazard domain or crisis management domain from a limited perspective or for a particular type.
For example, Humanitarian eXchange Language (HXL) \cite{clark2015feasibility} and EDXL-RESCUER Ontology \cite{barros2015edxl} are mainly concerned with help and rescue aspect of crisis management domain.
However, the broader and more diverse nature of this domain requires a \textbf{comprehensive} and \textbf{abstract} modeling and representation.
Another deficiency is related to lack of incorporating relations into the conceptualization.
Thus, they indeed should be called vocabulary or taxonomy rather than ontology (e.g., Management of Crisis vocabulary (MOAC)).
These shortages motivated us to introduce an ontology that 
takes into consideration the prior conceptualizations (taxonomies and vocabularies) while relying on a promoted representation.
Also, it brings new concepts and relations which were ignored previously although playing an essential role in capturing situational awareness (e.g., surveillance information, human sensing report, humanitarian event: prayer, concepts synonyms and instances).    
This paper presents our contribution in providing an ontology for Emergency Managing and PlAnning abouT Hazard crIses (\textbf{empathi}). 

The rest of the paper is organized as follows: Section \ref{sec:vocab} presents the relevant state-of-the-art hazard or crisis vocabularies and compare their main features.
Section \ref{sec:integration} lists the external vocabularies which are integrated empathi.
Thereafter, Section \ref{sec:coreconcepts} introduces empathi along with its major top concepts.
Section \ref{sec:relatedwork} reviews the related work. We close with the conclusion and the future work.
\vspace{-0.4em}
\section{State-of-the-art Vocabularies}
\label{sec:vocab}
In this section, we present an overview of the state-of-the-art vocabularies concerned with hazard domain as well as crisis management domain. 
With this respect, Table \ref{tab:comptable} represents a succinct comparative study of these vocabularies. 
The first column (i.e., \textbf{Vocabulary}) states the name of the state-of-the-art vocabulary or ontology and \emph{empathi} is included in the last row (based on recency). The second column (i.e., \textbf{Domain Coverage}) mentions the particular areas of the hazard-related domain or crisis management domain which it covers. The third column (i.e., \textbf{URI}) checks whether or not the URI of the vocabulary is dereferenceable.
The fourth column (i.e., \textbf{F for File}) specifies the available format of the vocabulary (i.e., OWL, RDF, TTL).
The fifth column (i.e., \textbf{D for Documentation}) shows whether the vocabulary has online documentation or not.
Then, we list the significant publications utilizing this vocabulary within the sixth column (i.e., \textbf{MC for Major Citations}).
The seventh and eight columns represent the number of classes \textbf{\#C})  and relations \textbf{\#R}) specified within the vocabulary.
The last column represents the external resources (i.e., \textbf{IV for Imported Vocabularies}) imported by their respective vocabulary in the first column. In the following, we shortly describe each vocabulary.

\begin{table*}[hpbt]
\centering
\scriptsize
\begin{tabular}{ |l|l|c|
c|c|c|l|l|l|l|}
 \hline
 \textbf{Vocabulary} & \textbf{Domain Coverage} & \textbf{U} & \textbf{F} & \textbf{D}  & \textbf{MC} & \textbf{\#C} & \textbf{\#R} & \textbf{IV} \\
 \hline
 \multirow{2}{*}{HXL}    & Disaster,Geography,Damage, & \cmark  & TTL & \cmark  & \cite{liu2013ontologies,clark2015feasibility,poveda2015designing,} & 50 & 66 & \cite{battle2011geosparql, brickley2007foaf, miles2005skos} \\ 
& Organization, Humanitarian Response  &    &  &   & \cite{ceballosusing,zavarella2014ontology,shah2013sierra, moi2016design,burel2017dores} & & &  \\\hline

\multirow{4}{*}{MOAC}  & Impact of Crisis, Recovery    & \xmark   & RDF  & \cmark  &\cite{liu2013ontologies,shih2013democratizing} & 70 & 30 & \cite{brickley2007foaf,morrow2011independent} \\ 
& and Response Activities,  &    &  &   & \cite{zavarella2014ontology,aid2017context} & & & \\
& Geo-locations   &    &  &   & \cite{case2015integration,roman2017infrarisk} & & & \\ \hline
 
\multirow{3}{*}{SMEM}  & Social Media and Emergency   & \xmark    & \xmark & \cmark  & - & - & - & \cite{jonkman2005global, lee2012ontology, nalchigar2010ontology, carroll2005named, brickley2007foaf, miles2005skos} \\
& Management  & & & & & & &  \cite{ breslin2005towards, brickley2003w3c, clark2015feasibility, van2011design, kim2008social, shotton2010cito} \\\hline

\multirow{4}{*}{DO} & Temporal and Spatial   & \cmark & Web-  & \xmark  & - &97 & - & - \\ 
&  Concepts, Impact, &    &App &   & & & & \\
& Rehabilitation and Facilities &   & &   & & & & \\
& Facilities &   & &   & & & & \\ \hline

\multirow{1}{*}{ERO} & Report Specification &\xmark  & \xmark  & \cmark   &\cite{barros122015edxl, bitencourt2015emergencyfire,simas2017data, snaprud2016better, villela2013rescuer} & - & - & \cite{fernandez1997methontology, prieto2003faceted, gruninger1995methodology} \\\hline

\multirow{2}{*}{DoRES} & Events and Reports   & \cmark    & RDF & \cmark  & - &96 & 261 & \cite{brickley2007foaf, miles2005skos}  \\
& Specification     &    &  &   & & & &  \cite{breslin2005towards, wick2015geonames} \\ \hline

\multirow{2}{*}{EF} &Fire Disaster Specification,    &\xmark    &\xmark  &\cmark   & - &37 &90 & - \\
& Protocol Design and Planning    &    &  &   & & & & \\ \hline
\multirow{2}{*}{empathi} & Hazard Situational Awareness, &\cmark    &OWL  &\cmark   & - &423 & 338 & \cite{anderson2002federal, jonkman2005global, lee2012ontology, miles2005skos, brickley2007foaf} \\
& Crisis Management, Hazard Events&    &  &   & & & & \cite{wick2015geonames, nalchigar2010ontology, breslin2005towards, purohit2014identifying, bhatt2014assisting}  \\\hline
\end{tabular}
\caption{Comparison of the state-of-the-art Hazard-related Vocabularies, Taxonomy, and Ontologies. U, F, D respectively stand for referenceability of URI, availability of File and Documentation. Furthermore,  MC, C, R and IV respectively stand for Major Citations,  number of classes, number of relations, and imported vocabularies. ERO : EDXL-RESCUER Ontology, DO : Disaster Ontology, EF: Emergency Fire.}
\label{tab:comptable}

\end{table*}


\paragraph{\textbf{HXL}} HXL stands for Humanitarian eXchange Language. HXL\footnote{\url{https://github.com/hxl-team/HXL-Vocab/blob/master/Tools/hxl.ttl}} is a standard aiming at information sharing during humanitarian calamity by overcoming the burden of interoperability. HXL ontology has a total of 50 classes and 66 relations. Main concepts contained in HXL are \texttt{Place}, \texttt{Survey and assessment}, \texttt{Operation}, \texttt{Cash and Finance}, \texttt{Crisis}. 
Furthermore, HXL provides links to UN OCHA vocabularies such as Global Coordination Groups\footnote{\url{https://goo.gl/CD6vHY}}, Disaster Types\footnote{\url{https://reliefweb.int/taxonomy-descriptions\#disastertype}}, Organization Types\footnote{\url{https://goo.gl/Uzy9UA}}, Vulnerable groups and Humanitarian themes\footnote{\url{http://vocabulary.unocha.org}}. Furthermore, HXL provides a hashtag schema containing related social media tags such as \#channel, \#crisis, \#impact, \#event, etc. \cite{clark2015feasibility}.

\begin{figure}
    \centering
    \includegraphics[width=0.5\textwidth]{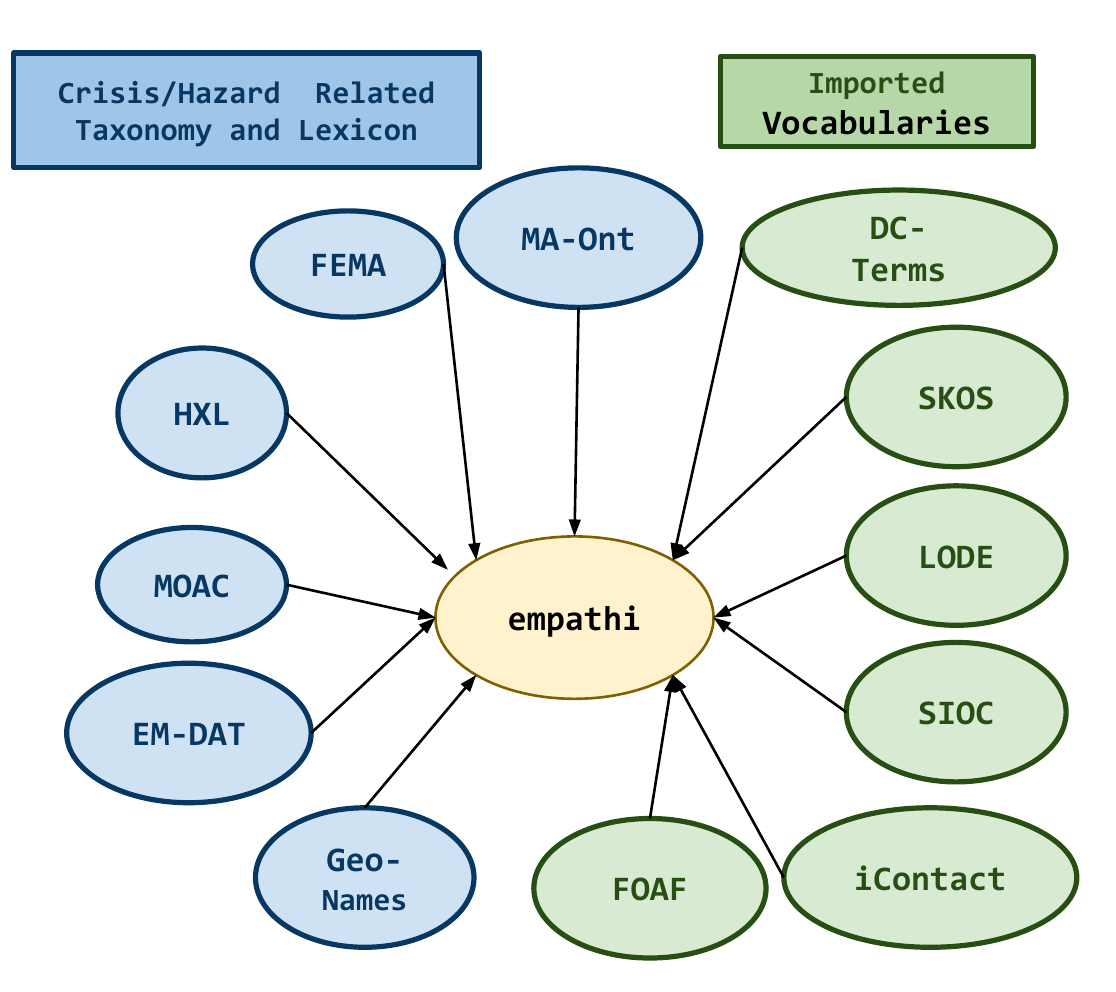}
   
    \caption{Importing existing ontologies in \textit{empathi}}
   
    \label{fig:integration}
\end{figure}

\paragraph{\textbf{MOAC}}  MOAC\footnote{\url{http://observedchange.com/moac/ns}} which is concerned with management of crisis is a vocabulary \cite{limbu2014management} providing concepts mainly related to crisis management. It was created by the Inter-Agency Standing Committee (IASC)\footnote{\url{https://goo.gl/ESn99F}}, Emergency Shelter Cluster in Haiti\footnote{\url{https://goo.gl/nDoa9F}}, UN-OCHA 3W Who What Where Contact Database\footnote{\url{https://goo.gl/jQLnYh}} and Ushahidi Platform\footnote{\url{https://goo.gl/XPSnyG}}. 

\paragraph{\textbf{DoRES}} DoRES stands for DOcument-Report-Event-Situation Ontology. 
DoRES\footnote{\url{https://goo.gl/Sw4XGt}} is an ontology sharing information between individuals and organizations using situational reports for describing the situation \cite{burel2017dores}.
It helps humanitarian organizations to structure their plans. 

\paragraph{\textbf{EDXL-RESCUER Ontology}}
EDXL-RESCUER\footnote{\url{http://www.rescuer-project.org}} stands for Emergency Data Exchange Language 
Reliable and Smart Crowdsourcing Solution for Emergency and Crisis Management. An ontology-based on EDXL developed for coordinating and interchanging information with the legacy system \cite{villela2013rescuer,barros2015edxl}.

\paragraph{\textbf{Emergency Fire (EF)}} 
It is an ontology explicitly designed for fires in the building. It comprises of 131 terms along with definitions created after subjective research. It serves as a protocol for information sharing, analysis, evaluation and comprehension by an organization in the situation of disaster caused by fire \cite{Bitencourt2015EmergencyFireAO}.

\paragraph{\textbf{Social Media Emergency Management (SMEM)}} During an unprecedented onset of a natural hazard, social media overflows with textual content about situational information, prayers, weather information, the impact of crisis and events. Of all the available information on the social media, which portion is contextually relevant and action-oriented to personnel in charge of a relief-giver organization. SMEM ontology provides concepts in a hierarchical structure which transforms high-volume of messy content to low-volume of action-oriented information \cite{moi2016ontology}.

\paragraph{\textbf{Disaster Ontology}}
It is one of the ontologies listed in Finnish Ontology Library Service ONKI\footnote{\url{https://onki.fi/en/}}. 
Disaster Ontology (DO)\footnote{\url{http://onki.fi/en/browser/overview/disaster}} is comprised of 97 concepts (classes) concerning man-made and natural hazard. This ontology is useful for managing disaster situations but disregards concepts related to social media (e.g., news reports, modality of data, surveillance, prayer and monitoring the status of the services provided by organizations).

\section{Integration of External Vocabularies}
\label{sec:integration}
In this section, we list the external vocabularies which partially integrated into empathi.
Not all of them are necessarily related to the domain of hazard or crisis management (we reuse generic concepts from well-known vocabularies, e.g., FOAF). 
Figure \ref{fig:integration} concisely represents an integration aims at reusing the existing vocabularies following ontology design methodologies (Methontology \cite{fernandez1997methontology} and NeOn \cite{suarez2012neon}) or interlinking \emph{empathi} to other vocabularies which enhance its visibility.

\begin{itemize}[label=-]
    \item \textbf{Federal Emergency Management Agency (FEMA)\footnote{\url{https://goo.gl/QtKzev}}} provides a glossary of terms related to disaster preparation and management \cite{anderson2002federal}.

\item \textbf{Emergency Disasters Database (EM-DAT)}\footnote{\url{http://www.emdat.be/Glossary. It is a database}} provides a precise definition of concepts and furthermore a categorization of disturbance-related events \cite{jonkman2005global}. 

\item \textbf{MA-Ont}\footnote{https://www.w3.org/ns/ma-ont} supports detailed properties describing media files and appropriate metadata mapping \cite{lee2012ontology}. 

\item \textbf{iContact}\footnote{ \url{http://ontology.eil.utoronto.ca/icontact.html}} provides conceptual classes for defining international addresses. It is relevant for using GeoNames \cite{wick2015geonames,nalchigar2010ontology} ontology for describing places.

\item \textbf{Friend Of A Friend (FOAF)}\footnote{\url{http://xmlns.com/foaf/spec/}} was created for describing people, relations, and associated events. 
Coupled with SIOC \cite{breslin2005towards} and disaster domain model \cite{bhatt2014assisting,purohit2014identifying}, it can describe social media communities formed during the disaster scenarios \cite{brickley2007foaf}.

\item \textbf{GeoNames}\footnote{\url{http://www.geonames.org/ontology/documentation.html}} is a part of GeoNames Database providing information about 11 million places (toponyms) covering all the countries. Integration of GeoNames ontology to our ontology adds geospatial semantic information which is critical for actionable hazard response. The ontology contains 150 classes and two relations forming 758 axioms on location dereferencing. Mapping syntax provided by this ontology is compatible with schema.org, DBpedia ontology, LinkedGeoData ontology and INSEE ontology \cite{wick2015geonames}. 

\item \textbf{Linked Open Descriptions of Events (LODE)\footnote{\url{http://linkedevents.org/ontology/}}}
defines event as an action which takes place at a certain time and has a specific location. It can be a historical action as well as a scheduled action. 
Thus, it provides the generic concept of \texttt{Event} along with locational (i.e., \texttt{atPlace}), temporal (i.e., \texttt{atTime}) aspects and people who play a role (i.e., \texttt{involvedAgent}).

\item \textbf{Simple Knowledge Organization System (SKOS)}\footnote{\url{https://www.w3.org/2004/02/skos/}}. We utilized this data model to describe the concepts of our domain. It provides a better organization of domain knowledge (i.e.,  Hazard Crisis) \cite{miles2005skos}.

\item \textbf{Semantically-Interlinked Online Communities (SIOC)}\footnote{\url{http://rdfs.org/sioc/spec/}} is a W3C ontological standard to describe information from online communities.  
It can support a volunteer or caregiver with actionable information in the realm of social media \cite{breslin2005towards}. 
\end{itemize}
\vspace{-1em}

\begin{table*}
   \centering
    \scriptsize
    \begin{tabular}{|l|l|l|} \hline
         \textbf{Hazard Type} & \textbf{Tweet}& \textbf{Hazard Concept}      \\ \hline
         \multirow{8}{*}{Flood}
         & 1. 188 killed, Airport closed they say that the runway came peeling off. it may take & Impact (AP, ID)
         \\ 
         & time to resume. &\\
         & 2. @? Hope people get adequate relief and no one is left out.Nation stands with& Event (HP)\\ 
         & Chennai. & \\
         & 3. @? in who can offer places to stay, pls fill the form for volunteers collating info & Event (VS)  \\
         & \#chennairains  &  \\
         & 4. @? Chennai has been declared disaster zone. Army has been deployed. Army & Service (H) \\
         & Helpline - +XXX XXX XXXX \#ChennaiRains & \\\hline
         \multirow{8}{*}{Hurricane}
         & 5. Surveillance video captures several people looting a Houston store  \#Harvey in & Event (CA)  \\
         & the wake of Hurricane &  \\
         & 6. The aftermath. 10,000 people now homeless because of Hurricane Irma. & Impact (AP), Event(CC)\\
         & \#WednesdayWisdom \#climatechange & \\
         & 7. \#HurricaneIrmaRecovery Drive for \#Homestead \& \#FloridaKeys today! Drop off & Service (S)\\
         & supplies at @? \#DJLMS \#dontgivebackjustgive. & \\
         & 8. At least 56 of Florida's 639 nursing homes still have *no* electricity this morning,  & Impact (ID)\\
         &  five days after \#HurricaneIrma & \\ \hline
         
         \multirow{7}{*}{Blizzard}
          & 9. @? "2-3 days, it could take before airlines begin to clear the backlog? @? at on & Impact (ID)\\
          & \#blizzard2016. & \\
          & 10. @?: I-75 in Kentucky closed due to large number of accidents, state patrol says  & Impact (AP,ID) \\
          &  \#blizzard2016 & \\
          & 11. @?: The baton is passed. Buoy 50 miles south of Wilmington \#Jonas \#blizzard2016   & Place (L)\\ 
          & 12. @?: Blizzard with "life and death implications" hits Washington, Mid-Atlantic & Place (L), Impact (AP) \\
          &  \#blizzard2016 & \\\hline
         
         \multirow{8}{*}{Landslide}
         & 13. @?: Be sure to follow @BGSLandslides for lots of up to date information on   & Report (ER) \\
         & landslides across the UK  \#StormFrank & \\
         & 14. \#StormFrank landslide at Glasscarraig Norman Motte \& Bailey in Co Wexford | @? & Place (L) \\
         &  \#archaeology \#floods  & \\
          & 15. SRI LANKA: At least 73 dead after week of flooding, landslides;  243,000 in temp & Service (SH), Impact (AP) \\
         & shelters | TorStar \#ExtremeWeather & \\
         & 16. @?: \#EcuadorEarthquake - landslides closing down roads \& making it challenging for & Impact (ID), Hazard Type(*) \\
         & help to reach hardest hit towns & \\\hline
    \end{tabular}
    \caption{ Sample of hazard-related tweets from different hazard types. AP: Affected Population, ID: Infrastructure Damage, HP: Human Prayer, VS: Volunteer Support, H: Helpline, CA: Criminal Activity, S: Supply, L: Location, ER: Expert Report, SH: Shelter, Hazard Type (*) : One hazard (Earthquake) causing another hazard (Landslide).}
    \label{table:tweetsamples}
\end{table*}

\section{Core Concepts of empathi}
\label{sec:coreconcepts}

\begin{figure}[!htbp] 
  \begin{center}
    \includegraphics[width=0.30\textwidth, trim=3.0cm 4.5cm 3.0cm 4.5cm, angle=-90]{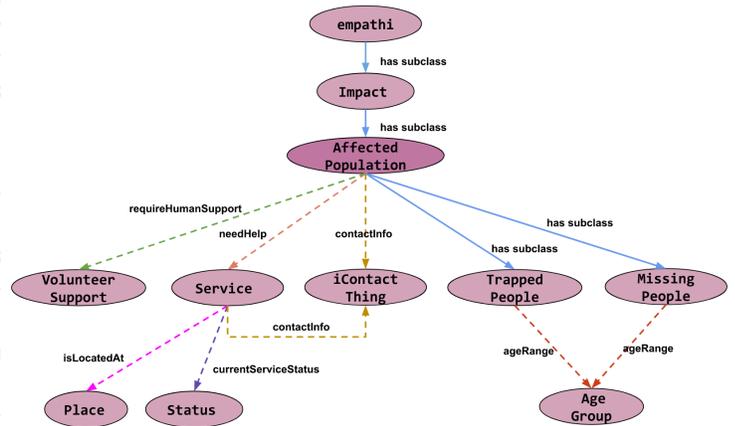}
  \end{center}
  \caption{Partial representation of the concept Affected Population in \emph{empathi}}
  \label{fig:empathi_example}
\end{figure}



As Table \ref{tab:comptable} shows, \emph{empathi} contains 423 classes and 338 relations. In figure \ref{fig:empathi_example}, concepts linked to "Affected Population" via solid lines, forms structural concepts ("is-a"/ "subclass"), while the concepts linked via colored dotted line are semantically related concepts to "Affected Population". Furthermore, in figure \ref{fig:empathi_example}, the concept "iContactThing" is taken from the vocabulary \emph{iContact} shown in figure \ref{fig:integration}. In the following, we present only the super-classes which imply the generic coverage.

\begin{itemize}[label=-]
    \item \textbf{Age Group:} this class groups people based on their age similarity by providing the following sub-classes (i) Adolescent, (ii) Adult, (iii) Child and (iv) Infant.
    \item \textbf{Event:} this concept defines events along with spatial and temporal constraints 
    happening in any phase of hazard. This concept embodies the following sub-classes (i) Climate Change (ii) Criminal Activity (iii) Emergency Exercises (iv) Evacuation Plan (v) Humanitarian Event (vi) Recovery Plan (vii) Rescue Operation (viii) State Mitigation Plan (ix) Volunteer Support and (x) Early Warning.
    
    \item \textbf{Facility:} defines an amenity made accessible for a specific purpose. It attributes to following sub-classes (i) Communication (ii) Electricity (iii) Gas Facility (iv) Water Facility and (v) Education Resource.
    
    \item \textbf{Hazard Type:} lists different types of hazards that can affect human community. It is an entity type that embodies sub-classes (i) Airburst (ii) Coastal erosion (iii) Drought (iv) Earthquake (v) Explosion (vi) Fire (vii) Flood (viii) Hurricane (ix) Landslide (x) Sandstorm (xi) Storm (xii) Tornado (xiii) Toxic Radioactivity (xiv) Tsunami (xv) Volcano and (xvi) Winterstorm. 
    
    \item \textbf{Hazard Phase:} categorizes different activities carried out by various organizations before, during and after a catastrophic event into three sub-classes (i) During Hazard (ii) Pre-Hazard and (iii) Post-Hazard.  
    
    \item \textbf{Impact:} a forceful negative affect on someone or something in an unprecedented manner. This concept embodies sub-classes (i) Affected Population (ii) Animal Loss (iii) Health Issues (iv) Food Shortage (v) Financial Crisis (vi) Contamination (vii) Infrastructure Damage (viii) Severity. 
    
    \item \textbf{Involved Actors:} People or Organisation associated (negative or positive) with any catastrophic event. Sub-classes of this concept are (i) Organisation (ii) People. 
    
    \item \textbf{Modality of Data:} Information (raw, structured or semi-structured) conveyed or represented by a particular arrangement or sequence of text, audio, video or photos. Sub-classes included by this concept are (i) Audio (ii) Photo (iii) Text and (iv) Video.
    
    \item \textbf{Place:} a physical surrounding defined by longitude, latitude, and area, providing a relative position of the someone or something during a hazard situation. One sub-class of the Place is Location, described by longitude, latitude, and the area of the affected place. 
    
    
    
    \item \textbf{Report:} documents evidence of the destruction caused by the natural disaster. All the activities carried out by various governmental and non-governmental organizations (NGOs) are stated in the report. The report is a way to keep people vigilant. Sub-classes encompassed under the concept: Report is (i) Expert Report (ii) Human Sensing Report and (iii) Media Report. 
    
    \item \textbf{Service:} is an act of providing support to someone in a situation of distressing incidents. Core sub-classes of this concept are (i) Financial Care (ii) Healthcare Service (iii) Helpline (iv) Human Remains Management (v) Resource and Information Centre (vi) Supply (vii) Transportation and (viii) Prayer Location.
    
    \item \textbf{Status:} defines the state of services that are planned during the pre/in/post hazard phases. Associated sub-classes are (i) Available (ii) Offered (iii) Requested and (iv) Unavailable. 
    
    \item \textbf{Surveillance Information:} A systematic, ongoing collection, collation, and analysis of data and the timely dissemination of information to those who need to know so that action can be taken. The surveillance concept in the setting of natural disasters can help to identify the resulting health-related needs which in turn, will lead to the more rational and effective deployment of resources to affected populations.

\end{itemize}

\begin{figure}
    \centering
    \includegraphics[width=0.5\textwidth]{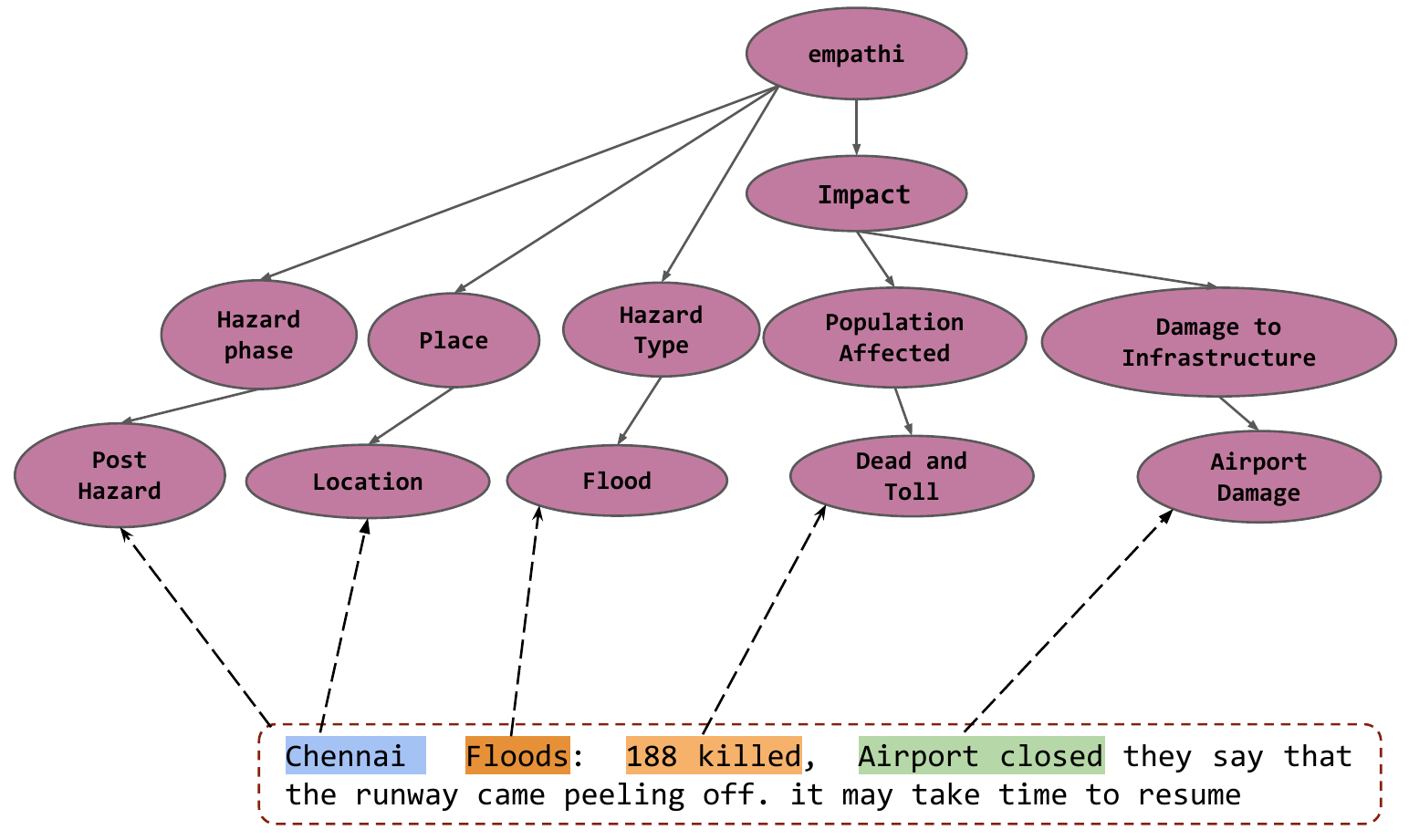}
   
    \caption{Mapping tweet and its words to \emph{empathi} concepts.}
   
    \label{fig:mapping}
\end{figure}

\vspace{-1.0em}
\section{Case Study}
\label{sec:casestudy}

A substantial impact of \emph{empathi} ontology is empowering us to annotate text semantically (e.g., tweets posted during hazard).
Figure \ref{fig:mapping} shows mapping of segments within a given tweet to \emph{empathi} concepts.
Thus, from an abstract level, this tweet is related to the flood occurred in Chennai and reports two associated impacts.
With this respect, in a first experiment, we compiled 53 million tweets from the 30 significant hazards happened in the past. Such as Hurricane Sandy in 2012 and Irma in 2017, Oklahoma Wildfires in 2017, Chennai Flood in 2005, Alaska Earthquake in 2018, Florida Rains in 2000 and 2016, Houston Floods in 2017, New-Zealand Earthquake in 2016, Typhoon Haima in 2016, Winter Storm Kayla in 2016, and many more.
After that, we identified the tweets related to the six central concepts of \emph{empathi} i.e., (i) impact, (ii) modality of data, (iii) hazard type, (iv) place, (v) transportation and (vi) surveillance.
Table \ref{tab:sub-concepts} and \ref{tab:concepts} show the statistics of identified tweets related to each of the chosen concepts and sub-concepts.
Furthermore, Table \ref{table:tweetsamples} represents samples of these tweets (i.e., column two) along with the mapped \emph{empathi} concepts (i.e., column three) from various hazard type (i.e., column one).

\begin{table}[!htbp]
    \centering
     \begin{tabular}{|l|c|}
            \hline
         \textbf{Sub-Concepts of empathi} & \textbf{\#Tweets } \\ \hline
         Water Facility (Fac.) & 218,968 \\ \hline
         Gas Facility (Fac.) & 33,047 \\ \hline
         Involved Organization (Inv.) & 4,249\\ \hline
         Severity (Imp.) & 1,344 \\ \hline
        \end{tabular}
        \caption{Mapping sub-concepts of \emph{empathi} to tweets of hazards. Fac.: Facility, Inv.: Involved, and Imp.: Impact are concepts}
        \label{tab:sub-concepts}
\end{table}

\begin{table}[]
    \centering
    \begin{tabular}{|l|c|}
         \hline
         \textbf{Concepts of empathi} & \textbf{\#Tweets } \\ \hline
         Hazard Type & 3,034,257 \\ \hline
         Impact & 618,446\\ \hline
         Modality of Data & 509,645 \\ \hline
         Facility & 258,117 \\ \hline
         Place & 16,397 \\ \hline
         Transportation & 6,694 \\ \hline
         Surveillance & 1,588 \\
         \hline 
        \end{tabular}
        \caption{Mapping concepts of \emph{empathi} to tweets of hazards.}
        \label{tab:concepts}
\end{table}

Extensive coverage by \emph{empathi} provides the capability of extracting structured information from unstructured and sparse content (e.g., Twitter) \cite{derczynski2013twitter}.  
For identifying relevant information from unstructured social media text, it is essential to map the words to ontology classes enable efficient classification of tweets as relevant and irrelevant to crisis domain. For instance in figure 2, the tweet: \textit{Chennai   Floods:  188 killed,  Airport closed they say that the runway came peeling off. it may take time to resume} is identified as a post-hazard tweet. \textit{Chennai} links to concept \textit{Place}, \textit{Floods} links to concept \textit{Hazard type}, \textit{188 killed} links to concept \textit{Affected population}, and \textit{Airport closed} links to concept \textit{Infrastructure Damage}. Moreover, such a procedure is termed as semantic annotation using expanded concepts (a.k.a. hypernyms).  It can improve understanding of social media messages which pose challenges like ill-formed sentences, ambiguous word senses,  poor syntactic structure, and implicit referencing. Semantic features formed using \emph{empathi} can enhance supervised and unsupervised learning in crisis domain \cite{khare2018classifying}.

\section{Evaluating Quality of \emph{empathi}}
To build up a quality ontology, we followed the principles of ontology methodologies such as NeON \cite{suarez2012neon} and Methontology \cite{fernandez1997methontology} which encourage the reuse of existing ontologies.
However, to quantitatively measure the quality of \emph{empathi}, we designed a user evaluation survey.
This survey contained 17 questions concerning with hierarchical, relational and lexical aspects of \emph{empathi}
inspired by  \cite{hlomani2014approaches}.
Precisely, the participants in the survey have to evaluate the following criteria:
(1) the correctness of structure (hierarchy)
(2) the correctness of relations between concepts, and
(3) lexical evaluation, i.e., quality of annotations associated with both concepts and relations.
In the following, we elaborate on these criteria.

\begin{figure}[!htbp]
\begin{subfigure}{0.5\textwidth}
    \includegraphics[width=0.95\textwidth, height=3.5cm ]{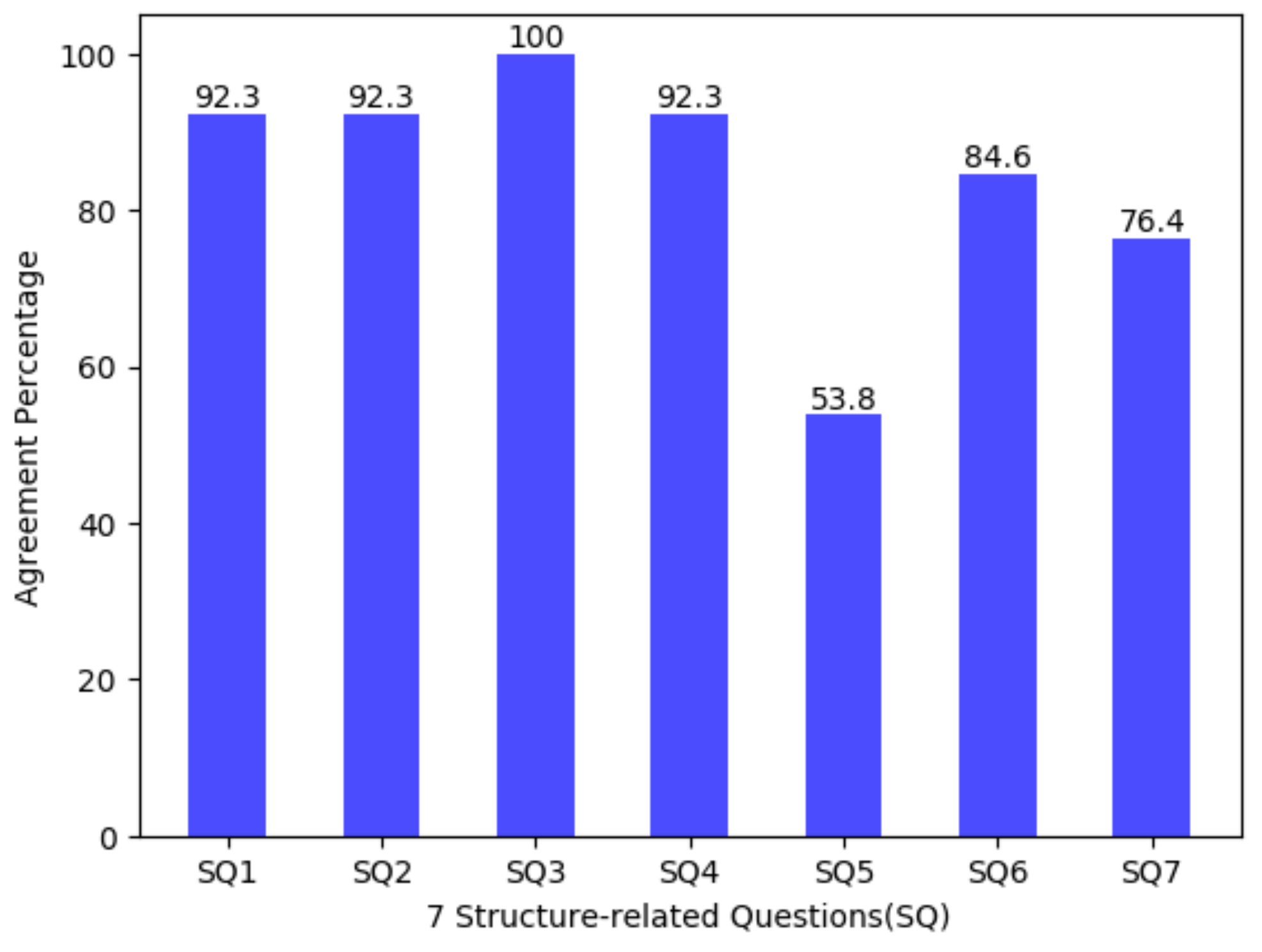}
    \caption{}
    \label{fig:SQ}
\end{subfigure}
\begin{subfigure}{0.5\textwidth}
    \includegraphics[width=0.9\textwidth, height=3.5cm ]{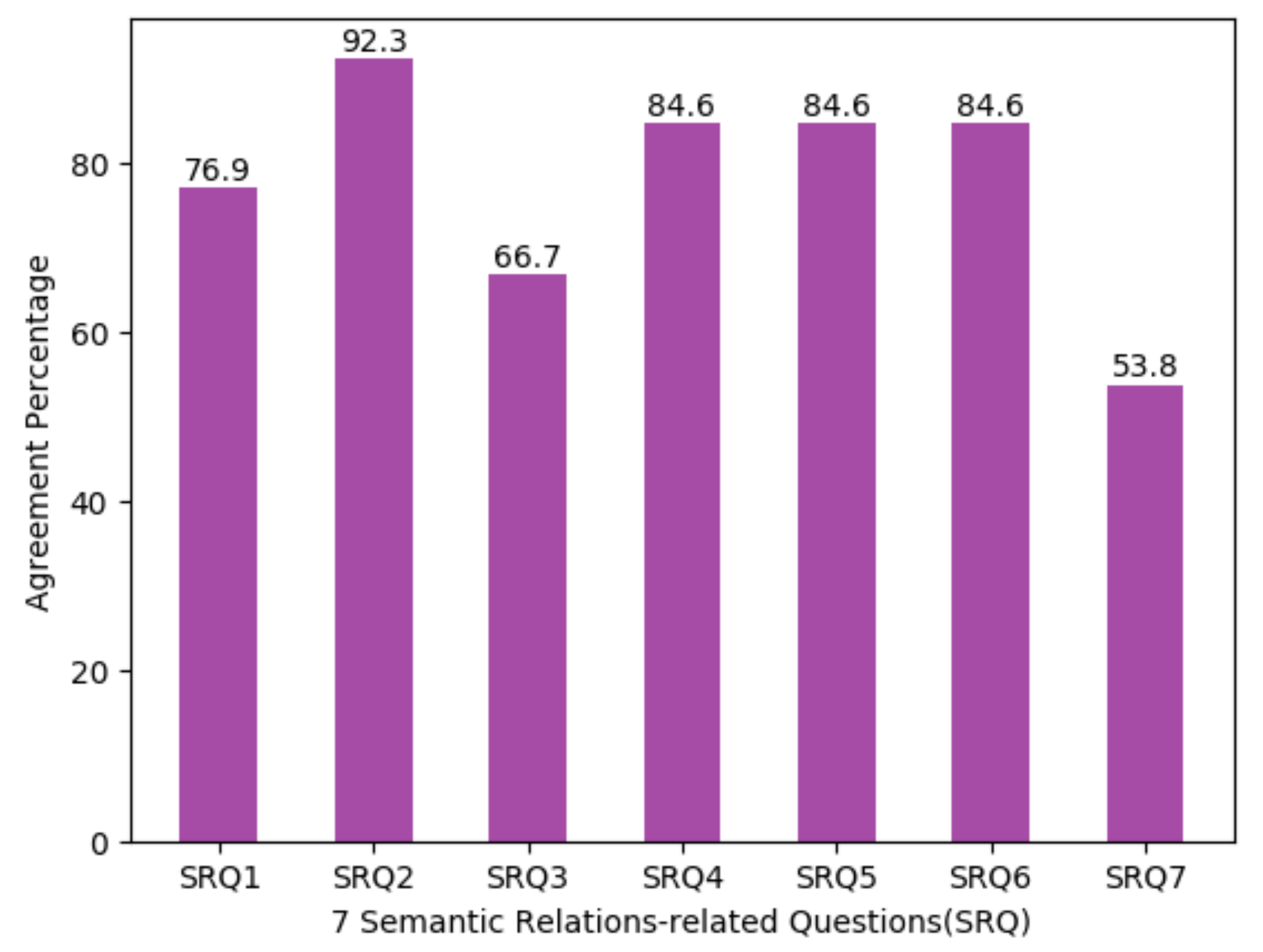}
    \caption{}
    \label{fig:SRQ}
\end{subfigure}
\begin{subfigure}{\textwidth}
    \includegraphics[width=0.5\textwidth, height=4.5cm]{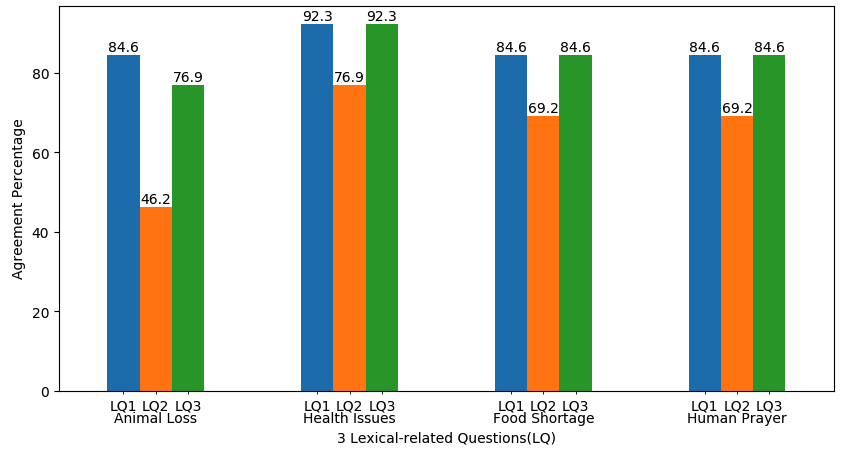}
    \caption{}
    \label{fig:LQ}
\end{subfigure}
\caption{Agreement of 13 evaluators on (a). 7 Questions that evaluate Structure of \emph{empathi}, (b). 7 Questions that evaluate Semantic Relations of \emph{empathi}, and (c). 3 Questions that evaluate Lexical characteristic of \emph{empathi} for 4 concepts: Animal Loss, Health Issues, Food Shortage, Human Prayer. Agreement percentage is calculated as the percentage of evaluators responded "agree" or "yes" in the ontology evaluation form.}
\label{fig:evaluation}
\end{figure}

\paragraph{\textbf{Structural evaluation}}
In this evaluation, the hierarchical structure is assessed concerning the correctness of "is-a" relationship as of whether the given concept A is-a particular type of the given concept B. For instance, ``Animal Loss'' is-a subclass of ``Impact'' in \emph{empathi}. Such evaluation is necessary to confirm the utility of ontology for classification task \cite{imran2015processing}. 
We presented parts of hierarchy in the survey, and asked the participants, how far this hierarchy makes sense to them.
They have to rate in the range 1 (fully disagree) -5 (fully agree).

\paragraph{\textbf{Semantic relational evaluation}}
Ontology is evaluated for holding semantically correct relations between concepts. For instance, there is a relation between the concept ``Affected Population'' and the concept ``Service'' referring "need for help". Thus, Affected Population is the domain and Service is the range. Concerning w.r.t. the prior survey \cite{imran2015processing}, having quality relations, higher capability for summarization, subgraph extraction, and contextualization tasks.
We represented a number of relations of \emph{empathi} to the participants and asked them whether or not they confirm having such a relation or not.

\paragraph{\textbf{Lexical evaluation}}
This part examines \emph{expressiveness}, \emph{completeness}, and \emph{clarity} of annotations of a given ontology. Expressiveness states the efficacy of the ontology to identify relevant information using natural language processing techniques. Completeness \cite{cordi2004checking} evaluates whether an illustration of a concept using definition and labels adequately define various scenarios in crisis domain. For instance, w.r.t. the given concept  ``Human Prayer'' the  definition ``prayer is a message to God from victim's relatives and family for protecting their lives and health" is provided along with the labels ``send prayers'', ``heart prayers'', ``heart praying'', ``join praying'', ``love prayers'', ``prayers affected'', ``temple'', ``church'', ``prayers city'', ``prayers families'', ``prayers involved'', etc. Clarity evaluates whether or not the concept name in the ontology is meaningful and easily understandable to human and machine. 
\paragraph{\textbf{Results}}
Our survey had 13 participants
and the results have been illustrated in figure \ref{fig:evaluation}. 

The structural evaluation section comprised of seven questions expressed as follows; (SQ1) Concerning a hazard situation, is ``Mental Stress" and ``Physical Stress" two important concepts under ``Health Issues''? (SQ2) Does ``No Effect", ``Minor", ``Major", ``Hazardous", ``Catastrophic" represents sub-classes of ``severity''? (SQ3)  Are ``Financial Crisis", ``Food Shortage", and ``Contamination" probable impacts of a Hazard? (SQ4) Are following triples meaningful: ``Animal Loss is-a Impact", ``Communication Lines Failure is-a Infrastructure Damage", ``Power Outage is-a Infrastructure Damage", and ``Survived People is-a Affected Population"? (SQ5) Do you consider non-government organization's (NGO) report is an expert report? (SQ6) Do health report, service feedback, and weather report define human sensing? (SQ7) Can ``News Agencies Report" and ``Social Media Report\footnote{\url{http://www.aidforum.org/topics/disaster-relief/top-100-twitter-accounts}}" be categorized under ``Media Report", a sub-class of ``Report" ? Questions SQ1-SQ6 were Yes/No questions, and SQ7 follows Likert Scale.
The detailed results are represented in Figure \ref{fig:SQ}. We observe the average agreement rate above 84.5\%.

Regarding evaluation of semantic relations, we designed seven questions as follows: SRQ1: Do you think the following triples make sense; ``Event occures in a Place", ``Service is offered in a Place", ``Each Hazard is associated to a couple of Event" and ``each Hazard leads to a couple of Services"? SRQ2: Can concepts, Event, and Service be linked to the concept ``Place” using ``isLocationAt" relation? SRQ3: Do you think two different types of Hazard can be related concerning Event, Service, and Place?  SRQ4: Is the ``currentStatus" relation between Facility and Status semantically correct? SRQ5: Is available, offered, requested and unavailable suitable categories for Status?  SRQ6: Can the concepts ``Service" and ``Organization" be concerning the concept ``Hazard Type"? SRQ7: Is the relation ``needHelp" correctly links ``Affected Population" with ``Organization" and ``Service"? Questions SRQ2-SRQ7 are Yes/No, and SRQ1 follow Likert Scale. From figure \ref{fig:SRQ}, an average agreement of 75.4\% was concluded in confirming above facts.

The lexical evaluation of \emph{empathi} was performed by representing definition and synonyms (or labels) describing each concept and asking participants to respond the following questions; (LQ1) Do labels appropriately represent the concept? (LQ2)  Are labels complete? (LQ3) Are definitions and labels enough clear? Figure \ref{fig:LQ} shows the results with the total agreement rate 78.8\%. 

\section{Related Work}
\label{sec:relatedwork}
The aftermath of the disaster causes agencies/organization to structure a plan for the thoughtful recovery of the area. An opportunity to hasten this process is a need for a knowledge structure containing concepts, constraints, and links that provide before-hand information for disaster preparedness and act. In \cite{fahland2008towards}, the author designed a process model of the aftermath of the disasters using the Petri-Net containing inter-linked concepts for efficiency after-disaster execution and adaptation. Furthermore, the cost associated with the recovery of an area urges the need to have a structured source of events, concepts, and relations that define a distressing incident. In \cite{murgante2009building}, the author defines the utility of an ontology providing seismic risk definition, prediction, and management to reduce damages. There is a recent study on crowdsourced emergency event detection in \cite{grolinger2013knowledge}. There, the authors propose the utilization of emerging knowledge from text using concepts and temporal information for events. In a recent work on crisis management \cite{benaben2017conceptual}, there has been a development of a suite of tools which can leverage our ontology for context-aware response generation. CrowdGeoKG \cite{chen2017crowdgeokg} is a framework using the entities in the OpenStreetMap and is enriched by Wikidata.

Events mainly characterize a disaster scenario, and it is essential to understand what is happening in an emergency situation. In \cite{song2015simulator}, a given user can utilize GPS data to create a simulation model for predicting human mobility. We assume that identification of core disaster domain-specific concepts can help in annotating the GPS data similar to in OpenStreetMap \cite{auer2009linkedgeodata}. 
In 2012, Hurrican Sandy brought in towering traffic on social sensing sites urging the need of an information filtering mechanism for assisting crisis coordination. A psycholinguistic driven domain-dependent lexicon was created in \cite{purohit2013crisis} for assistive crisis response. Moreover, tweets on Hurricane Sandy 2012 identified ``power blackout'' as one of the implications of the disaster. To identify its associated repercussions, one need a structured domain-model. For instance, crashing out of power affected the medical facility. Hence, there is a need for an ontology to bridge the facilities in emergency situation \cite{bhatt2014assisting}. In \cite{hampton2017constructing}, the author created a twitter stream of deceptive and peripheral messages using a knowledge source assisting Public Information Officers (PIOs) to make conscious decisions in an emergency situation.

Table 1 motivates creating \textit{empathi}. Prior ontologies related to crisis management and situational awareness fails to provide structural, lexical and relational benefits needed for extracting situation-specific information extraction from social media content \cite{imran2015processing}. It is either because these ontologies are diverse in their subject areas or are incomplete concerning referencing, documentation, and expressibility.  On analyzing these state-of-the-art vocabularies, we identified subject areas, vocabularies concerning each subject area and used OWL to provide a formal representation. Since ontologies can be domain-specific (crisis domain) or generic (e.g., FOAF), we incorporate concepts relevant in addressing various issues in crisis management. For instance; concept Location was taken from Geonames and iContact, crisis-related concepts from HXL and MOAC. Hence, we addressed all the queries raised in \cite{liu2013ontologies}.  
Human social communication during an emergency event provides real-time insight into various domains such as facility, events, impact, report, surveillance, organization involved and activities carried out during and after the hazard. Extracting actionable information from active social channels is challenges because of 2 reasons: (1) absence of an ontology that map multiple concepts, (2) completeness and expressiveness of the ontology.  We provide a utility based case-study (section \ref{sec:casestudy}) where we used our ontology for mapping 53M tweets to the concepts and sub-concepts in \emph{empathi}. Mapping of social media content to ontology concepts will improve classification and summarization task using state-of-the-art natural language processing and learning techniques.

\section{Conclusion and Future Work}
We propose \emph{empathi} ontology as a crisis domain archetype that aids crisis management, hazard situational awareness, and hazard events during emergency scenarios. In our study, we demonstrated the prowess of ontology by its integration with relevant and state-of-the-art crisis vocabularies. Moreover, its efficacy was assessed through appropriate evaluation of its quality across three dominant criteria: structure, lexical and semantic relations using the human judge. Furthermore, we illustrated its effectiveness concerning social media domain by linking tweet content to ontology concepts.  
The ontology has been created using semantic web language (OWL) and links to FOAF, SIOC, DC-terms, and LODE. 
We aim to extend the ontology in following directions; the first direction is to introduce Internet of Things (IoT) for disaster management. A second possible direction is to improve the ontology with additional ontology quality tools as recommended by the PerfectO methodology\footnote{\url{http://perfectsemanticweb.appspot.com/?p=ontologyValidation}}. For instance, use of LODE tool provides automatic documentation. WebVOWL tool provides an automated graph visualization. A third direction is to disseminate more the ontologies on ontology catalogs such as Linked Open Vocabularies (LOV) and LOV4IoT. LOV4IoT could be refined and extended to support the environment domain with various use cases such as flooding, fire, earthquake, tsunami.


\section*{Acknowledgment}
We are thankful to responders in providing their unbiased evaluation for ontology quality assessment. 
We acknowledge support from the National Science Foundation (NSF) award EAR 1520870: Hazards SEES: Social and Physical
Sensing Enabled Decision Support for Disaster Management and Response. Any opinions, findings, and conclusions/recommendations expressed in this material are those of the author(s) and do not necessarily reflect the views of the NSF.


\small{
\bibliographystyle{IEEEtran}
\bibliography{reference}}

\begin{thebibliography}{10}
\providecommand{\url}[1]{#1}
\csname url@samestyle\endcsname
\providecommand{\newblock}{\relax}
\providecommand{\bibinfo}[2]{#2}
\providecommand{\BIBentrySTDinterwordspacing}{\spaceskip=0pt\relax}
\providecommand{\BIBentryALTinterwordstretchfactor}{4}
\providecommand{\BIBentryALTinterwordspacing}{\spaceskip=\fontdimen2\font plus
\BIBentryALTinterwordstretchfactor\fontdimen3\font minus
  \fontdimen4\font\relax}
\providecommand{\BIBforeignlanguage}[2]{{%
\expandafter\ifx\csname l@#1\endcsname\relax
\typeout{** WARNING: IEEEtran.bst: No hyphenation pattern has been}%
\typeout{** loaded for the language `#1'. Using the pattern for}%
\typeout{** the default language instead.}%
\else
\language=\csname l@#1\endcsname
\fi
#2}}
\providecommand{\BIBdecl}{\relax}
\BIBdecl

\bibitem{clark2015feasibility}
T.~Clark, C.~Ke{\ss}ler, and H.~Purohit, ``Feasibility of information
  interoperability in the humanitarian domain,'' in \emph{AAAI Symposium},
  2015.

\bibitem{barros2015edxl}
R.~Barros, P.~Kislansky, L.~Salvador, R.~Almeida, M.~Breyer, and L.~G. Pedraza,
  ``Edxl-rescuer ontology: Conceptual model for semantic integration.'' in
  \emph{ISCRAM}, 2015.

\bibitem{liu2013ontologies}
S.~Liu, C.~Brewster, and D.~Shaw, ``Ontologies for crisis management: A review
  of state of the art in ontology design and usability.'' in \emph{ISCRAM},
  2013.

\bibitem{poveda2015designing}
G.~Poveda, E.~Serrano, and M.~Garijo, ``Designing emergency management services
  by ontology driven social simulation,'' \emph{IT CoNvergence PRActice}, 2015.

\bibitem{}
N.~Sugiura, Y.~Shigeta, N.~Fukuta, N.~Izumi, and T.~Yamaguchi, ``Towards
  on-the-fly ontology construction--focusing on ontology quality improvement,''
  \emph{The Semantic Web: Research and Applications}, pp. 1--15, 2004.

\bibitem{battle2011geosparql}
R.~Battle and D.~Kolas, ``Geosparql: enabling a geospatial semantic web,''
  \emph{Semantic Web Journal}, 2011.

\bibitem{brickley2007foaf}
D.~Brickley and L.~Miller, ``Foaf vocabulary specification 0.91,'' 2007.

\bibitem{miles2005skos}
A.~Miles, B.~Matthews, M.~Wilson, and D.~Brickley, ``Skos core: simple
  knowledge organisation for the web,'' in \emph{International Conference on
  Dublin Core and Metadata Applications}, 2005.

\bibitem{ceballosusing}
H.~G. Ceballos, E.~Olivarria, and E.~A. Fernandez, ``Using hxl hashtags for
  semantic annotation of egovernment tabular data.''

\bibitem{zavarella2014ontology}
V.~Zavarella, H.~Tanev, R.~Steinberger, and E.~Van~der Goot, ``An
  ontology-based approach to social media mining for crisis management.'' in
  \emph{SSA-SMILE@ ESWC}, 2014.

\bibitem{shah2013sierra}
S.~M. Shah, C.~Brewster, and D.~Shaw, ``Sierra: cooperative request-response
  for resource management in disasters using semantic web principles,'' 2013.

\bibitem{moi2016design}
M.~Moi and N.~Rodehutskors, ``Design of an ontology for the use of social media
  in emergency management,'' in \emph{International Conferences ICT, WBC,
  BIGDACI and TPMC}, 2016.

\bibitem{burel2017dores}
G.~Burel, L.~S. Piccolo, K.~Meesters, and H.~Alani, ``Dores—a three-tier
  ontology for modelling crises in the digital age,'' in \emph{ISCRAM}, 2017.

\bibitem{shih2013democratizing}
F.~Shih, O.~Seneviratne, I.~Liccardi, E.~Patton, P.~Meier, and C.~Castillo,
  ``Democratizing mobile app development for disaster management,'' in
  \emph{Joint Proceedings of the Workshop on AI Problems and Approaches for
  Intelligent Environments and Workshop on Semantic Cities}, 2013.

\bibitem{morrow2011independent}
N.~Morrow, N.~Mock, A.~Papendieck, and N.~Kocmich, ``Independent evaluation of
  the ushahidi haiti project,'' \emph{Development Information Systems
  International}, 2011.

\bibitem{aid2017context}
A.~Aid and I.~Rassoul, ``Context-aware framework to support situation-awareness
  for disaster management,'' \emph{International Journal of Ad Hoc and
  Ubiquitous Computing}, 2017.

\bibitem{case2015integration}
P.~O. C.~U. CASE, ``Integration of provenance-enabled crowdsourced information
  with traditional disaster management information using linked open data,''
  2015.

\bibitem{roman2017infrarisk}
D.~Roman, D.~Sukhobok, N.~Nikolov, B.~Elves{\ae}ter, and A.~Pultier, ``The
  infrarisk ontology: Enabling semantic interoperability for critical
  infrastructures at risk from natural hazards,'' in \emph{OTM}.\hskip 1em plus
  0.5em minus 0.4em\relax Springer, 2017.

\bibitem{jonkman2005global}
S.~N. Jonkman, ``Global perspectives on loss of human life caused by floods,''
  \emph{Natural hazards}, 2005.

\bibitem{lee2012ontology}
W.~Lee, W.~Bailer, T.~B{\"u}rger, P.-A. Champin, J.-P. Evain, V.~Malais{\'e},
  T.~Michel, F.~Sasaki, J.~S{\"o}derberg, F.~Stegmaier \emph{et~al.},
  ``Ontology for media resources 1.0,'' \emph{W3C recommendation}, 2012.

\bibitem{nalchigar2010ontology}
S.~Nalchigar and M.~S. Fox, ``An ontology for open 311 data,'' 2010.

\bibitem{carroll2005named}
J.~J. Carroll, C.~Bizer, P.~Hayes, and P.~Stickler, ``Named graphs,'' \emph{Web
  Semantics: Science, Services and Agents on the World Wide Web}, 2005.

\bibitem{breslin2005towards}
J.~G. Breslin, A.~Harth, U.~Bojars, and S.~Decker, ``Towards
  semantically-interlinked online communities,'' in \emph{ESWC}, 2005.

\bibitem{brickley2003w3c}
D.~Brickley, ``W3c basic geo vocabulary,'' 2003.

\bibitem{van2011design}
W.~R. Van~Hage, V.~Malais{\'e}, R.~Segers, L.~Hollink, and G.~Schreiber,
  ``Design and use of the simple event model (sem),'' \emph{Web Semantics:
  Science, Services and Agents on the World Wide Web}, 2011.

\bibitem{kim2008social}
H.-L. Kim, J.~G. Breslin, S.-K. Yang, and H.-G. Kim, ``Social semantic cloud of
  tag: Semantic model for social tagging,'' in \emph{KES International
  Symposium on Agent and Multi-Agent Systems: Technologies and Applications},
  2008.

\bibitem{shotton2010cito}
D.~Shotton, ``Cito, the citation typing ontology,'' in \emph{Journal of
  biomedical semantics}, 2010.

\bibitem{barros122015edxl}
R.~Barros12, P.~Kislansky, L.~Salvador12, R.~Almeida, M.~Breyer, L.~G. Pedraza,
  and V.~Vieira12, ``Edxl-rescuer ontology: an update based on faceted taxonomy
  approach,'' 2015.

\bibitem{bitencourt2015emergencyfire}
K.~Bitencourt, F.~Dur{\~a}o, and M.~Mendon{\c{c}}a, ``Emergencyfire: An
  ontology for fire emergency situations,'' in \emph{Proceedings of the 21st
  Brazilian Symposium on Multimedia and the Web}, 2015.

\bibitem{simas2017data}
F.~Simas, R.~Barros, L.~Salvador, M.~Weber, and S.~Amorim, ``A data exchange
  tool based on ontology for emergency response systems,'' in \emph{Research
  Conference on Metadata and Semantics Research}, 2017.

\bibitem{snaprud2016better}
M.~Snaprud, J.~Radianti, and D.~Svindseth, ``Better access to terminology for
  crisis communications,'' in \emph{International Conference on Information
  Technology in Disaster Risk Reduction}, 2016.

\bibitem{villela2013rescuer}
K.~Villela, V.~Vieira, M.~Mendon{\c{c}}a, J.~Torres, and S.~Graffy,
  ``Rescuer-dow-reliable and smart crowdsourcing solution for emergency and
  crisis management,'' \emph{Munich, Germany}, 2013.

\bibitem{fernandez1997methontology}
M.~Fern{\'a}ndez-L{\'o}pez, A.~G{\'o}mez-P{\'e}rez, and N.~Juristo,
  ``Methontology: from ontological art towards ontological engineering,'' 1997.

\bibitem{prieto2003faceted}
R.~Prieto-D{\'\i}az, ``A faceted approach to building ontologies,'' in
  \emph{IEEE IRI}, 2003.

\bibitem{gruninger1995methodology}
M.~Gr{\"u}ninger and M.~S. Fox, ``Methodology for the design and evaluation of
  ontologies,'' 1995.

\bibitem{wick2015geonames}
M.~Wick, B.~Vatant, and B.~Christophe, ``Geonames ontology,'' \emph{URL
  http://www. geonames. org/ontology}, 2015.

\bibitem{anderson2002federal}
C.~Anderson, \emph{The Federal Emergency Management Agency (FEMA)}, 2002.

\bibitem{purohit2014identifying}
H.~Purohit, A.~Hampton, S.~Bhatt, V.~L. Shalin, A.~P. Sheth, and J.~M. Flach,
  ``Identifying seekers and suppliers in social media communities to support
  crisis coordination,'' \emph{CSCW}, 2014.

\bibitem{bhatt2014assisting}
S.~P. Bhatt, H.~Purohit, A.~Hampton, V.~Shalin, A.~Sheth, and J.~Flach,
  ``Assisting coordination during crisis: a domain ontology based approach to
  infer resource needs from tweets,'' in \emph{Web science}, 2014.

\bibitem{limbu2014management}
M.~Limbu, ``Management of a crisis (moac) vocabulary specification.
  observedchange, january 2012,'' 2014.

\bibitem{Bitencourt2015EmergencyFireAO}
K.~Bitencourt, F.~Dur{\~a}o, and M.~G. Mendonça, ``Emergencyfire: An ontology
  for fire emergency situations,'' in \emph{WebMedia}, 2015.

\bibitem{moi2016ontology}
M.~Moi, N.~Rodehutskors, and R.~Koch, ``An ontology for the use of quality
  evaluated social media data in emergencies,'' \emph{IADIS International
  Journal on WWW/Internet}, 2016.

\bibitem{suarez2012neon}
M.~C. Suarez-Figueroa, A.~Gomez-Perez, and M.~Fernandez-Lopez, ``{The NeOn
  Methodology for Ontology Engineering},'' in \emph{Ontology engineering in a
  networked world}.\hskip 1em plus 0.5em minus 0.4em\relax Springer, 2012.

\bibitem{derczynski2013twitter}
L.~Derczynski, A.~Ritter, S.~Clark, and K.~Bontcheva, ``Twitter part-of-speech
  tagging for all: Overcoming sparse and noisy data,'' in \emph{Proceedings of
  the International Conference Recent Advances in Natural Language Processing
  RANLP 2013}, 2013.

\bibitem{khare2018classifying}
P.~Khare, G.~Burel, and H.~Alani, ``Classifying crises-information relevancy
  with semantics,'' 2018.

\bibitem{hlomani2014approaches}
H.~Hlomani and D.~Stacey, ``Approaches, methods, metrics, measures, and
  subjectivity in ontology evaluation: A survey,'' \emph{Semantic Web Journal},
  vol.~1, no.~5, pp. 1--11, 2014.

\bibitem{imran2015processing}
M.~Imran, C.~Castillo, F.~Diaz, and S.~Vieweg, ``Processing social media
  messages in mass emergency: A survey,'' \emph{ACM Computing Surveys (CSUR)},
  2015.

\bibitem{cordi2004checking}
V.~Cord{\i} and V.~Mascardi, ``Checking the completeness of ontologies: a case
  study from the semantic web,'' in \emph{Proc. of the CILC’04 Workshop},
  2004.

\bibitem{fahland2008towards}
D.~Fahland and H.~Woith, ``Towards process models for disaster response,'' in
  \emph{International Conference on Business Process Management}, 2008.

\bibitem{murgante2009building}
B.~Murgante, G.~Scardaccione, and G.~Las~Casas, ``Building ontologies for
  disaster management: seismic risk domain,'' \emph{Urban and Regional Data
  Management}, 2009.

\bibitem{grolinger2013knowledge}
K.~Grolinger, M.~A. Capretz, E.~Mezghani, and E.~Exposito, ``Knowledge as a
  service framework for disaster data management,'' in \emph{IEEE WETICE},
  2013.

\bibitem{benaben2017conceptual}
F.~Benaben, A.~Montarnal, S.~Truptil, M.~Lauras, A.~Fertier, N.~Salatge, and
  S.~Rebiere, ``A conceptual framework and a suite of tools to support crisis
  management,'' 2017.

\bibitem{chen2017crowdgeokg}
J.~Chen, S.~Deng, and H.~Chen, ``Crowdgeokg: Crowdsourced geo-knowledge
  graph,'' in \emph{China Conference on Knowledge Graph and Semantic
  Computing}, 2017.

\bibitem{song2015simulator}
X.~Song, Q.~Zhang, Y.~Sekimoto, R.~Shibasaki, N.~J. Yuan, and X.~Xie, ``A
  simulator of human emergency mobility following disasters: Knowledge transfer
  from big disaster data.'' in \emph{AAAI}, 2015.

\bibitem{auer2009linkedgeodata}
S.~Auer, J.~Lehmann, and S.~Hellmann, ``Linkedgeodata: Adding a spatial
  dimension to the web of data,'' in \emph{ISWC}, 2009.

\bibitem{purohit2013crisis}
H.~Purohit, ``Crisis response coordination in online communities,'' 2013.

\bibitem{hampton2017constructing}
A.~J. Hampton, S.~Bhatt, A.~Smith, J.~Brunn, H.~Purohit, V.~L. Shalin, J.~M.
  Flach, and A.~P. Sheth, ``Constructing synthetic social media stimuli for an
  emergency preparedness functional exercise,'' 2017.

\end{thebibliography}



%

\end{document}